\documentclass[titlepage,12pt]{article}
\usepackage{amssymb,psfig,epsfig,pslatex}
\usepackage{cite}
\makeatletter
\def\@sect#1#2#3#4#5#6[#7]#8{\ifnum #2>\c@secnumdepth
  \def\@svsec{}\else
  \refstepcounter{#1}\edef\@svsec{\csname the#1\endcsname.\hskip0.5em}\fi
  \@tempskipa #5\relax
  \ifdim \@tempskipa>\z@
    \begingroup
      #6\relax
      \@hangfrom{\hskip #3\relax\@svsec}{\interlinepenalty \@M #8\par}%
    \endgroup
    \csname #1mark\endcsname{#7}\addcontentsline
      {toc}{#1}{\ifnum #2>\c@secnumdepth \else
        \protect\numberline{\csname the#1\endcsname}\fi #7}%
  \else
    \def\@svsechd{#6\hskip #3\@svsec #8\csname #1mark\endcsname
      {#7}\addcontentsline{toc}{#1}{\ifnum #2>\c@secnumdepth \else
        \protect\numberline{\csname the#1\endcsname}\fi #7}}%
  \fi \@xsect{#5}}
\@addtoreset{equation}{section}
\makeatother
\renewcommand\thesection{\Roman{section}}
\renewcommand\theequation{\ifnum \value{section}>0
 \thesection.\arabic{equation}%
\else
\arabic{equation}%
\fi}

\newcommand{\nn}{\nonumber}

\def\Li2{{{\rm Li}_2}}

\renewcommand{\thefootnote}{\small\fnsymbol{footnote}}

\begin{document}
\begin{titlepage}
\noindent 
DESY 05-043  \hfill \\
SDU-HEP200501 \hfill \\
\vspace{0.01cm}
\begin{center}
{\LARGE {\bf Top quark pair production and decay  \\ 
at a polarized photon collider}  \\
\vspace{2cm}
\large{\bf 
A. Brandenburg\,$^{a,}$\footnote{work supported  
by a Heisenberg fellowship
of DFG.}, Z. G. Si\,$^{b,}$\footnote{work supported
in part by NSFC and NCET}}}
\par\vspace{1cm}
$^a$DESY-Theorie, 22603 Hamburg, Germany\\
$^b$Department of Physics, Shandong University, Jinan, Shandong
250100, China
\par\vspace{1cm}
{\bf Abstract:}\\
\parbox[t]{\textwidth}
{
Top quark pair production 
by (polarized) $\gamma\gamma$ collisions offers 
an interesting  testing ground of the Standard Model and its extensions.
In this Letter we present results for 
differential cross sections of top quark pair production
and decay including QCD radiative corrections. We take into account
the full dependence on the top quark spins. We give analytic and numerical
results for single and double differential 
angular distributions of $t\bar{t}$ decay
products which are due to top quark polarizations and spin correlations
in the intermediate state.   
}
\end{center}
\vspace*{2cm}

\end{titlepage}
%
%
\setcounter{footnote}{0}
\renewcommand{\thefootnote}{\arabic{footnote}}
\setcounter{page}{1}

\section{Introduction} 

At a future linear lepton collider, backscattered laser light may provide
very high-energy photons \cite{Ginzburg:1982yr}, which  would 
allow for a very interesting physics program
\cite{Brodsky:1994nf,Badelek:2001xb}.
In particular,  top quark pair production is possible 
with large rates in (un)polarized photon photon fusion. 
The measurement of
the process $\gamma \gamma \rightarrow t\bar{t}X$ is an important test of the
Standard Model (SM).
The first order QCD corrections to this process 
have already been calculated and found to be 
large\cite{Drees:1992eh,Kuhn:1992fx,Jikia:1996bi,Jikia:2000rk}.
The electroweak virtual plus soft-photonic ${\cal O}(\alpha)$
corrections are also known \cite{Denner:1995ar}.
This process will 
also provide information on  possible anomalous $\gamma t\bar{t}$ 
couplings \cite{Choi:1995kp,Grzadkowski:2003tf} 
without contributions from $Z t\bar{t}$
couplings present in $e^+e^-$ collisions.

Once the Higgs
boson is discovered, it will be of primary importance to determine whether
its properties are as predicted within the SM.
In this respect, the process
$\gamma\gamma \rightarrow t\bar{t}X$ may play an important role.
For example, heavy quark production in polarized 
$\gamma\gamma$ collisions will help to determine the parity of the 
Higgs boson produced as a resonance and decaying into top quark pairs 
\cite{Jikia:1996bi,Jikia:2000rk}. 
In particular, if a Higgs boson is no CP eigenstate, spin correlations
of the top quark pairs will help to probe the scalar and pseudoscalar
couplings of the Higgs boson to the top quark  
\cite{Anlauf:1995mu,Bernreuther:1997af}. 

For this kind of studies, predictions for top quark pair
production and decay 
at a photon collider must be as precise as possible within the SM.
In particular, the spin state of the intermediate $t\bar{t}$ pair 
must be taken into account. (The role of the top quark polarization
in probing the $t\bar{t}$ threshold dynamics in $\gamma\gamma$ collisions
was discussed in \cite{Fadin:1994pj}).

The purpose of this paper is therefore to study the processes 
\begin{equation}
\gamma \gamma \rightarrow t \bar{t} X \rightarrow \ell + \ell'+X
, \quad \ell+{\mbox{ \ jets}}+X, \quad{\mbox {\ all\ jets}},
\end{equation}
where $\ell$ stands for a charged lepton, 
with polarized photons from backscattered laser beams.
We include QCD radiative corrections and take
into account polarization and  spin correlation effects of the
intermediate $t\bar{t}$ pairs. 

Leading order results and QCD corrections for the cross section and
for top quark spin observables in the process 
$\gamma\gamma\to t\bar{t}X$ are summarized
in sections \ref{LO} and \ref{parton level}.
Numerical results to order $\alpha^2\alpha_s$ 
for the effective lepton collider cross section and for several decay
distributions are given in section \ref{nummerical results}.


\section{Kinematics and leading order results}
\label{LO}
 
The production of top quark pairs by
photon scattering at leading order $\alpha^2\alpha_s^0$ is
described by the reaction 
\begin{equation}\label{bornreac}
\gamma(p_1,\lambda_1)+\gamma(p_2,\lambda_2)\to t(k_1,s_t)
+\bar{t}(k_2,s_{\bar{t}}).
\end{equation}
Here, $p_1,p_2,k_1$ and $k_2$ denote the momenta of the particles,
$\lambda_1 $ and $\lambda_2$ are the helicities of the photons, 
and the vectors $s_t$ and $s_{\bar{t}}$ describe the spins of top quark 
and antiquark. These fulfil the relations
\begin{equation}
s_t^2=s_{\bar{t}}^2=-1\qquad {\mbox{and}}\qquad k_1\cdot s_t
=k_{\bar{t}}\cdot s_{\bar{t}}=0.
\end{equation}
In the (anti)top rest frame the spin of the (anti)top is described
by a unit vector $\hat{\bf s}_{t}$ ($\hat{\bf s}_{\bar{t}}$). We choose
the specific rest frames that are obtained by a rotation-free Lorentz
boost from the zero momentum frame of the $t\bar{t}$ quarks 
($t\bar{t}$-ZMF). Both the $t\bar{t}$-ZMF and the $t$ and $\bar{t}$ rest 
frames will be used to construct spin observables from the final
state momenta of the $t\bar{t}$ decay products.
We use the $t\bar{t}$-ZMF rather than the c.m. frame of the colliding
high-energy photons, since the latter system is probably 
more difficult to reconstruct experimentally. For the $2\to 2$ process of
Eq.~(\ref{bornreac}), the two frames coincide. 

The differential cross section for the process of Eq.~(\ref{bornreac})  
can be written as follows:
  \begin{equation}
d\sigma(\lambda_1,\lambda_2,s_t,s_{\bar{t}})={N\over 2 s_{\gamma\gamma}}
|M_0|^2d\Gamma_2,
\end{equation}
where  the two-particle phase space measure is denoted by $d\Gamma_2$,
$s_{\gamma\gamma}=(p_1+p_2)^2$ and $N=3$ is the number of colours.
A simple calculation gives:
\begin{eqnarray}
|M_0|^2&=&{16\alpha^2Q_t^4\pi^2\over (1-\beta^2z^2)^2}
\Bigg\{
A_0+B_0 \left[p_1\cdot (s_t+s_{\bar{t}})\right]+
B_0|_{\lambda_1\leftrightarrow\lambda_2}
\left[p_2\cdot (s_t+s_{\bar{t}})\right]
\nonumber \\ 
&+&C_0(s_t\cdot s_{\bar{t}})
+ D_0(p_1\cdot s_t)(p_2\cdot s_{\bar{t}})
+D_0|_{z\rightarrow -z}(p_2\cdot s_t)(p_1\cdot s_{\bar{t}})
\Bigg\}
\end{eqnarray}
with  
\begin{eqnarray}
A_0&=& 1+2\beta^2(1-z^2)-\beta^4\left[1+(1-z^2)^2
\right]+\lambda_1\lambda_2
\left[
1-2\beta^2(1-z^2)-\beta^4z^2(2-z^2)
\right],\\
B_0 &=& {4m\over s_{\gamma\gamma}}\left[
\lambda_1(1-2\beta^2+\beta^2z^2)+\lambda_2(1-\beta^2z^2)
\right],\\
C_0&=& 1-2\beta^2+\beta^4\left[1+(1-z^2)^2\right]+\lambda_1\lambda_2
\left[
1-2\beta^2+\beta^4z^2(2-z^2)
\right],\\
D_0&=& -{4(1+\beta z)(1-z^2)(1-\lambda_1\lambda_2)\beta^2\over 
s_{\gamma\gamma}}.
\end{eqnarray}
Here, $Q_t=2/3$, $m$ is the top quark mass,
\begin{equation}
\beta=\sqrt{1-{4m^2\over s}},
\end{equation}
and $z$ is the cosine of the scattering angle in the $t\bar{t}$-ZMF, i.e.
$z=\hat{\bf p}_{\gamma}\cdot\hat{\bf k}$, where $\hat{\bf p}_\gamma$ 
($\hat{\bf k}$) is the direction of one of the photons 
(of the top quark) in that frame.

\section{NLO results for $\gamma\gamma\to t\bar{t}X$} 
\label{parton level}
In this section we present  results for the inclusive reaction
\begin{equation}
\gamma\gamma \to t\bar{t}X
\end{equation}
to order $\alpha^2\alpha_s$. 
Apart from the cross section we study observables
that depend on the spins of the top quark and antiquark.
For polarized photons, observables of the form
\begin{equation}\label{os}
O^s=2{\bf S}_{t}\cdot \hat{\bf a}
\end{equation}
can have non-zero expectation values. Here, $\hat{\bf a}$ is an arbitrary 
reference direction and ${\bf S}_{t}$ is the top quark 
spin operator.
The expectation value of $O^s$ is related to a single spin asymmetry:
\begin{equation}
\langle O^s \rangle = {\sigma(\uparrow)-\sigma(\downarrow)\over
\sigma(\uparrow)+\sigma(\downarrow)}, 
\end{equation}
where the arrows on the right-hand side refer to the spin state of
the top quark with respect to the quantization axis $\hat{\bf a}$.
We will consider here two choices for $\hat{\bf a}$, 
\begin{eqnarray}\label{adir}
\hat{\bf a}=\hat{\bf k} \qquad && {\rm (helicity \ basis)},\nonumber \\
\hat{\bf a}=\hat{\bf p} \qquad && {\rm (beam \ basis)},
\end{eqnarray}
where $\hat{\bf k}$ denotes the direction of the top quark in the
$t\bar{t}$-ZMF and $\hat{\bf p}$ is the direction
of the lepton beam  coming from the left in that frame, which coincides
to good approximation with the direction of one of the high-energy 
photons. Top quark polarization 
perpendicular to the plane spanned
by $\hat{\bf p}$ and $\hat{\bf k}$ is induced by absorptive parts
in the one-loop amplitude. This effect is, however, quite
small ($\sim$ a few percent) \cite{Bernreuther:1994vd}.  

Analogous observables may of course be defined for the top antiquark.

Apart from the above single spin observables, we also consider observables
of the form 
\begin{equation}\label{od}
O^d=4({\bf S}_{t}\cdot \hat{\bf a})({\bf S}_{\bar{t}}\cdot \hat{\bf b}).
\end{equation}
Here, $\hat{\bf a}$ and $\hat{\bf b}$ are arbitrary 
reference directions and ${\bf S}_{\bar{t}}$ is the top antiquark 
spin operator.
The expectation value of $O^d$ is related to a double spin asymmetry:
\begin{equation}
\langle O^d \rangle = {\sigma(\uparrow\uparrow)+\sigma(\downarrow\downarrow)
-\sigma(\uparrow\downarrow)-\sigma(\downarrow\uparrow)\over
\sigma(\uparrow\uparrow)+\sigma(\downarrow\downarrow)
+\sigma(\uparrow\downarrow)+\sigma(\downarrow\uparrow)}. 
\end{equation}
For the reference directions we will consider here 
\begin{eqnarray}\label{abdir}
&&\hat{\bf a}=-\hat{\bf b}=\hat{\bf k}, \nonumber \\
&&\hat{\bf a}=\hat{\bf b}=\hat{\bf p}.
\end{eqnarray}
Finally, we also present results for the observable
\begin{equation}\label{orot}
\tilde{O}^d={4\over 3}{\bf S}_{t}\cdot {\bf S}_{\bar{t}}.
\end{equation}
\par
The above double spin asymmetries $\langle O^d \rangle$
and $\langle \tilde{O}^d \rangle$ have also proved
useful for an analysis of  spin correlations
of top quark pairs in hadronic collisions \cite{Bernreuther:2004jv}.
\par
The NLO cross section for the reaction $\gamma\gamma\to t\bar{t}X$
may be written in terms of two scaling functions:
\begin{eqnarray}
\sigma(\hat{s},m,\lambda_1,\lambda_2)&=&{\alpha^2\*Q_t^4\over m^2}\*
\left[
c^{(0)}(\rho,\lambda_1,\lambda_2)
+4\*\pi\*\alpha_s\*c^{(1)}(\rho,\lambda_1,\lambda_2)
\right].
\end{eqnarray}
Likewise, the unnormalized expectation values of the above
spin observables are of the form 
\begin{eqnarray}
\sigma
\*\langle O_a\rangle 
&=&{\alpha^2\*Q_t^4\over m^2}\*
\left[
d^{(0)}_a(\rho,\lambda_1,\lambda_2)
+4\*\pi\*\alpha_s\*d^{(1)}_a(\rho,\lambda_1,\lambda_2)
\right],
\end{eqnarray}
where $a=1$ corresponds to the observable $\tilde{O}^d$ defined 
in Eq.~(\ref{orot}),
$a=2 (3)$ to the observable $O^d$ defined in Eq.~(\ref{od}) 
in the helicity (beam) basis,
and $a=4 (5)$ corresponds to 
the single spin observable $O^s$ defined in Eq.~(\ref{os})
in the helicity (beam) basis. 
The variable $\rho$ is defined as
\begin{equation}
\rho={4\*m^2\over s_{\gamma\gamma}}.
\end{equation}
The lowest order scaling functions $c^{(0)}$ and $d^{(0)}_a$
can be computed analytically. We use the following
auxiliary functions, which vanish in the limit $\beta=\sqrt{1-\rho}\to 0$:
\begin{eqnarray}
\ell_1&=&{1\over \beta}\left[\ln(x)+2\*\beta\right],\nonumber \\
\ell_2&=&{1\over \beta^3}\left[\ln(x)+2\*\beta+{2\over 3}\*\beta^3\right],
\nonumber \\ 
\ell_3&=&{1\over \beta^5}\left[\ln(x)+2\*\beta+{2\over 3}\*\beta^3+
{2\over 5}\*\beta^5\right],
\end{eqnarray}
where $x=(1-\beta)/(1+\beta)$.
We then obtain:
\begin{eqnarray}
c^{(0)}(\rho,\lambda_1,\lambda_2) &=& N\*\pi\*\beta\*\rho\*
\left\{1+\rho-\rho^2+\lambda_1\*\lambda_2-\left[1+\rho-{\rho^2\over 2}
-\lambda_1\*\lambda_2\right]\*\ell_1\right\},\nonumber \\
d^{(0)}_1(\rho,\lambda_1,\lambda_2) &=& -{N\*\pi\*\beta\*\rho\over 3}\*
\left\{1+\rho+\rho^2+(1+2\*\rho)\*\lambda_1\*\lambda_2
+\left[1-{\rho^2\over 2}
-(1+\rho)\*\lambda_1\*\lambda_2\right]\*\ell_1\right\},\nonumber \\
d^{(0)}_2(\rho,\lambda_1,\lambda_2) &=& N\*\pi\*\beta\*\rho\*
\Bigg\{{1+8\*\rho-7\*\rho^2+\rho^3+(5-3\*\rho+\rho^2)
\*\lambda_1\*\lambda_2\over 3} \nonumber \\ &-& 
\left[
-1+\rho-2\*\rho^2+{\rho^3\over 2}+\left(1+{\rho^2\over 2}
\right)\*\lambda_1\*\lambda_2
\right]\*\ell_2
\Bigg\},\nonumber \\
d^{(0)}_3(\rho,\lambda_1,\lambda_2) &=& N\*\pi\*\beta\*\rho\*
\Bigg\{
-{9-20\*\sqrt{\rho}-6\*\rho+14\*\rho^{3/2}+29\*\rho^2
+6\*\rho^{5/2}-20\*\rho^3+3\*\rho^4\over 15} \nonumber \\
&+&{-21-20\*\sqrt{\rho}+25\*\rho+14\*\rho^{3/2}-16\*\rho^2
+6\*\rho^{5/2}-3\*\rho^3\over 15}
\*\lambda_1\*\lambda_2 \nonumber \\
&+&\Bigg[-1-4\*\rho-\rho^2-2\*\sqrt{\rho}+{\rho^3\over 2}
-2\*\rho^{3/2}+\rho^{5/2}\nonumber \\
&+& \left(1+2\*\sqrt{\rho}+3\*\rho+{\rho^2\over 2}\right)\*\lambda_1\*\lambda_2
\Bigg]\*(1-\sqrt{\rho})^2\*\ell_3
\Bigg\},\nonumber \\
d^{(0)}_4(\rho,\lambda_1,\lambda_2) &=& N\*\pi\*\beta\*\rho\*
\left\{\sqrt{1-\rho}-{\rho\over 2}\*\ln(x)\right\}
\*\left(\lambda_1+\lambda_2\right),\nonumber \\
d^{(0)}_5(\rho,\lambda_1,\lambda_2) &=& -N\*\pi\*\beta\*\rho\*
\Bigg\{
{1-\rho\over 3}\*\left(1-6\*\sqrt{\rho}+\rho^{3/2}\right)
\nonumber \\
&+&{1-\sqrt{\rho}\over 2}\*\left(
2+2\*\sqrt{\rho}+3\*\rho-\rho^{3/2}-\rho^2
\right)\*\ell_2
\Bigg\}
\*\left(\lambda_1-\lambda_2\right).
\end{eqnarray}  
The functions $c^{(1)}$ and  $d^{(1)}_a$ are 
obtained by a numerical integration.  The scaling functions
for the spin-averaged cross section and all spin observables are plotted
in Figs.~1-3 for different choices of the photon helicities as a function
of 
\begin{equation}
\eta={1\over \rho}-1.
\end{equation}
\begin{figure}[t]
  \unitlength1.0cm
  \begin{center}
    \begin{picture}(4.5,4.5)
      \put(-6.65,-3){\psfig{figure=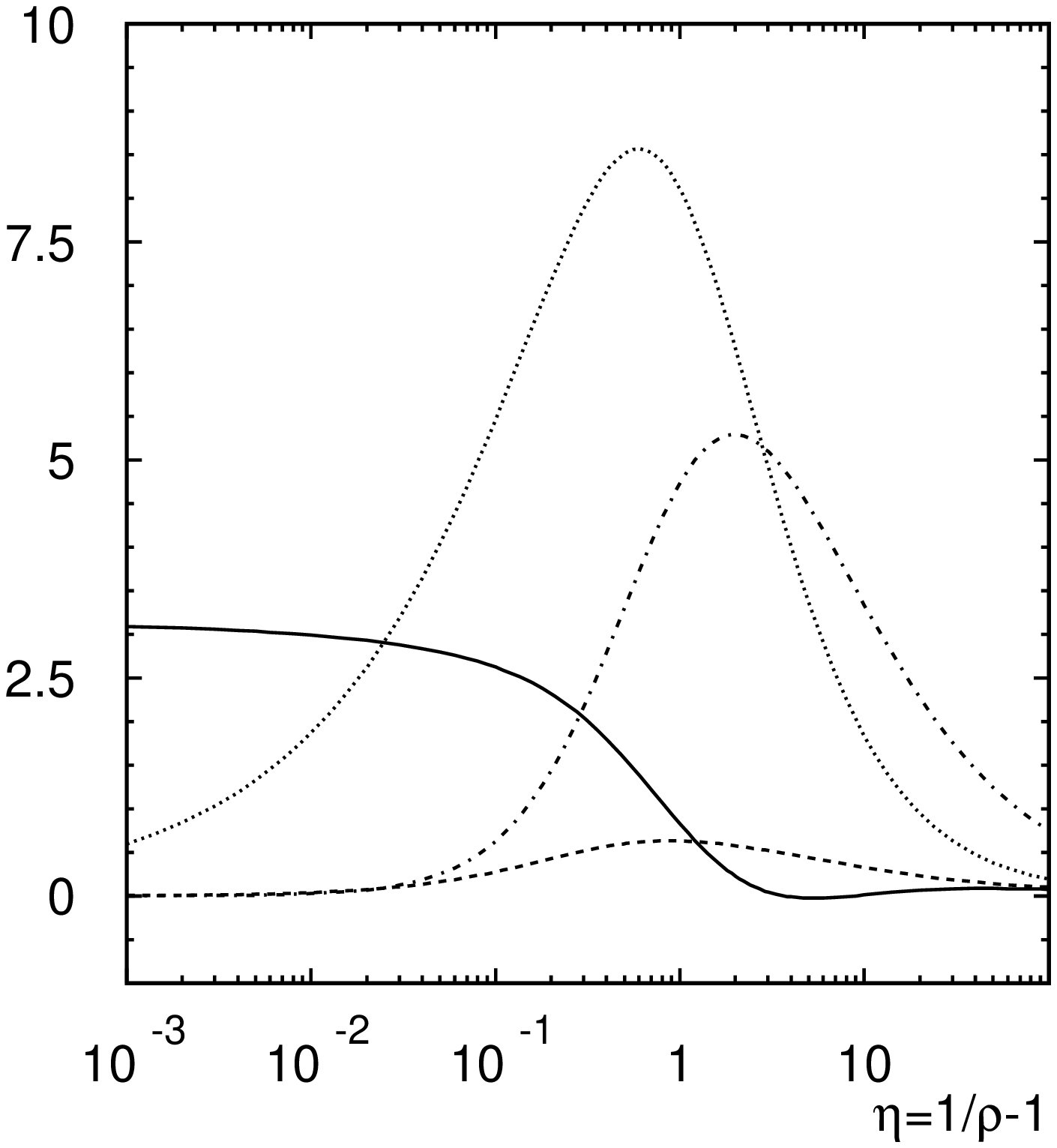,width=9cm}}
      \put(2.65,-3){\psfig{figure=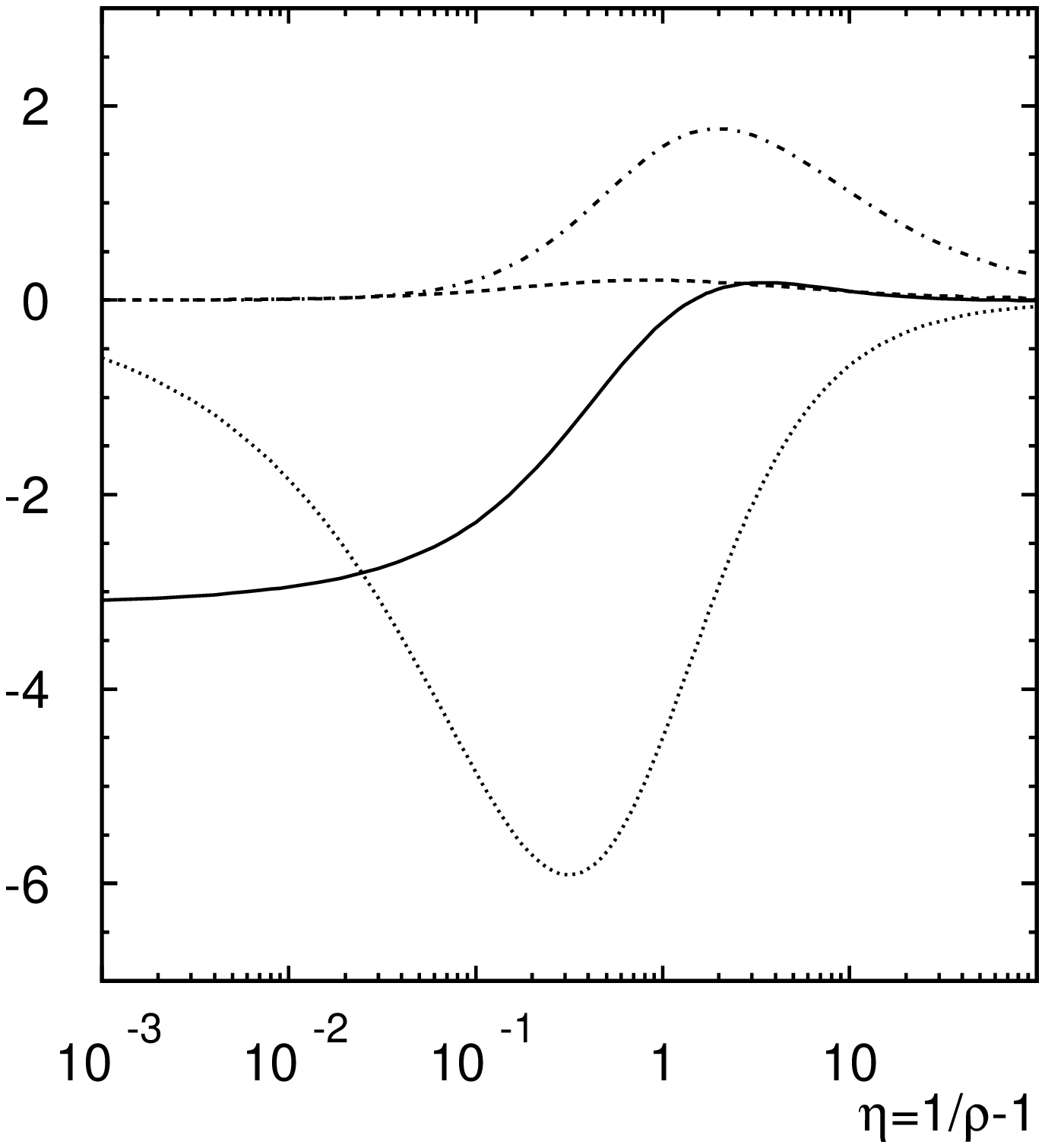,width=9cm}}
    \end{picture}
    \vskip 2.75cm
    \caption{\it Left: Scaling functions $c^{(0)}(\rho,1,1)$ (dotted),
  $c^{(0)}(\rho,1,-1)$ (dash-dotted), $c^{(1)}(\rho,1,1)$ (full),
  and $c^{(1)}(\rho,1,-1)$ (dashed).
   Right: Scaling functions $d^{(0)}_1(\rho,1,1)$ (dotted),
  $d^{(0)}_1(\rho,1,-1)$ (dash-dotted), $d^{(1)}_1(\rho,1,1)$ (full), 
  and $d^{(1)}_1(\rho,1,-1)$ (dashed).}
    \label{fig:o01}
  \end{center}
\end{figure}
\begin{figure}[h!]
  \unitlength1.0cm
  \begin{center}
    \begin{picture}(4.5,4.5)
      \put(-6.65,-3.5){\psfig{figure=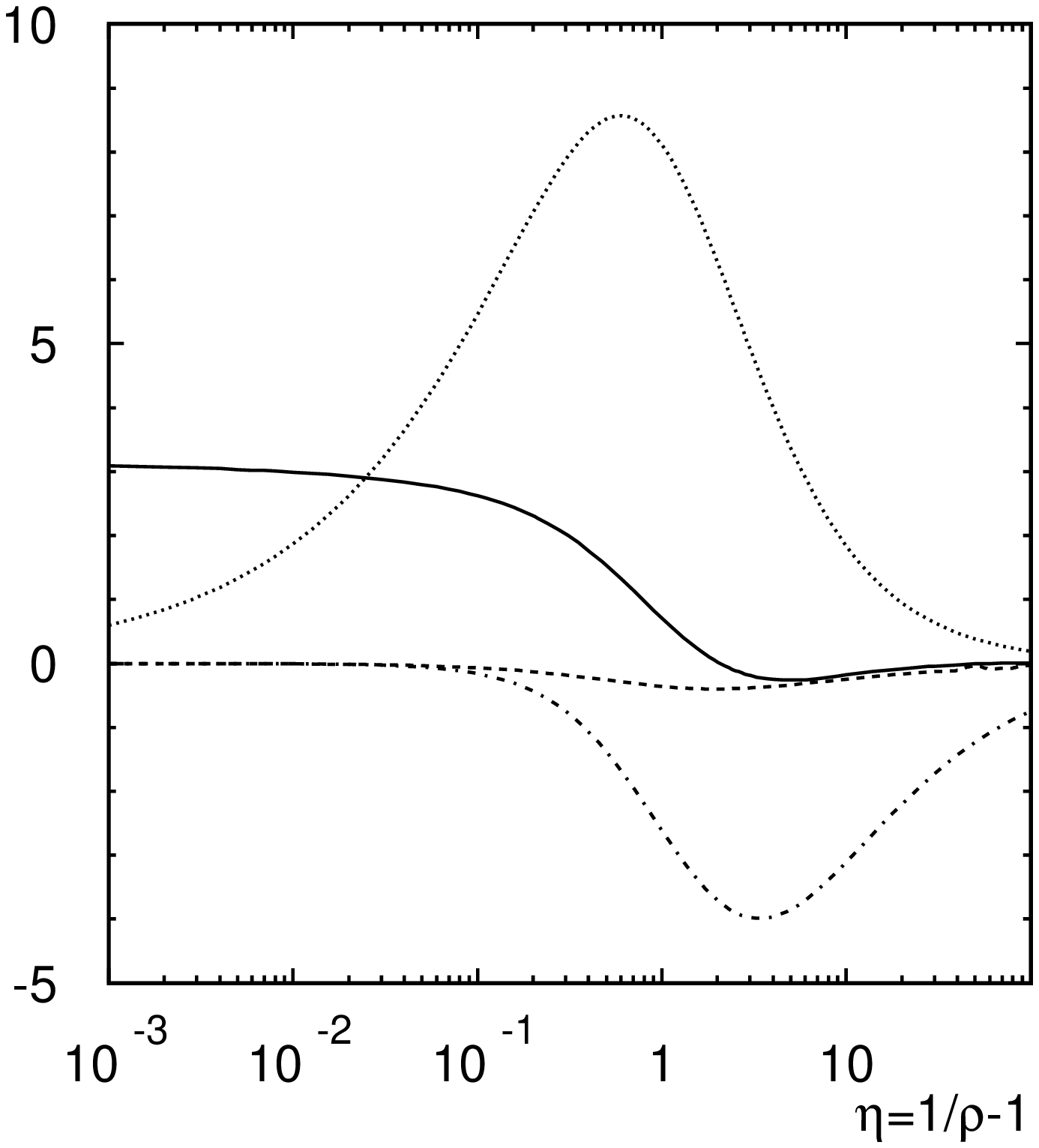,width=9cm}}
      \put(2.65,-3.5){\psfig{figure=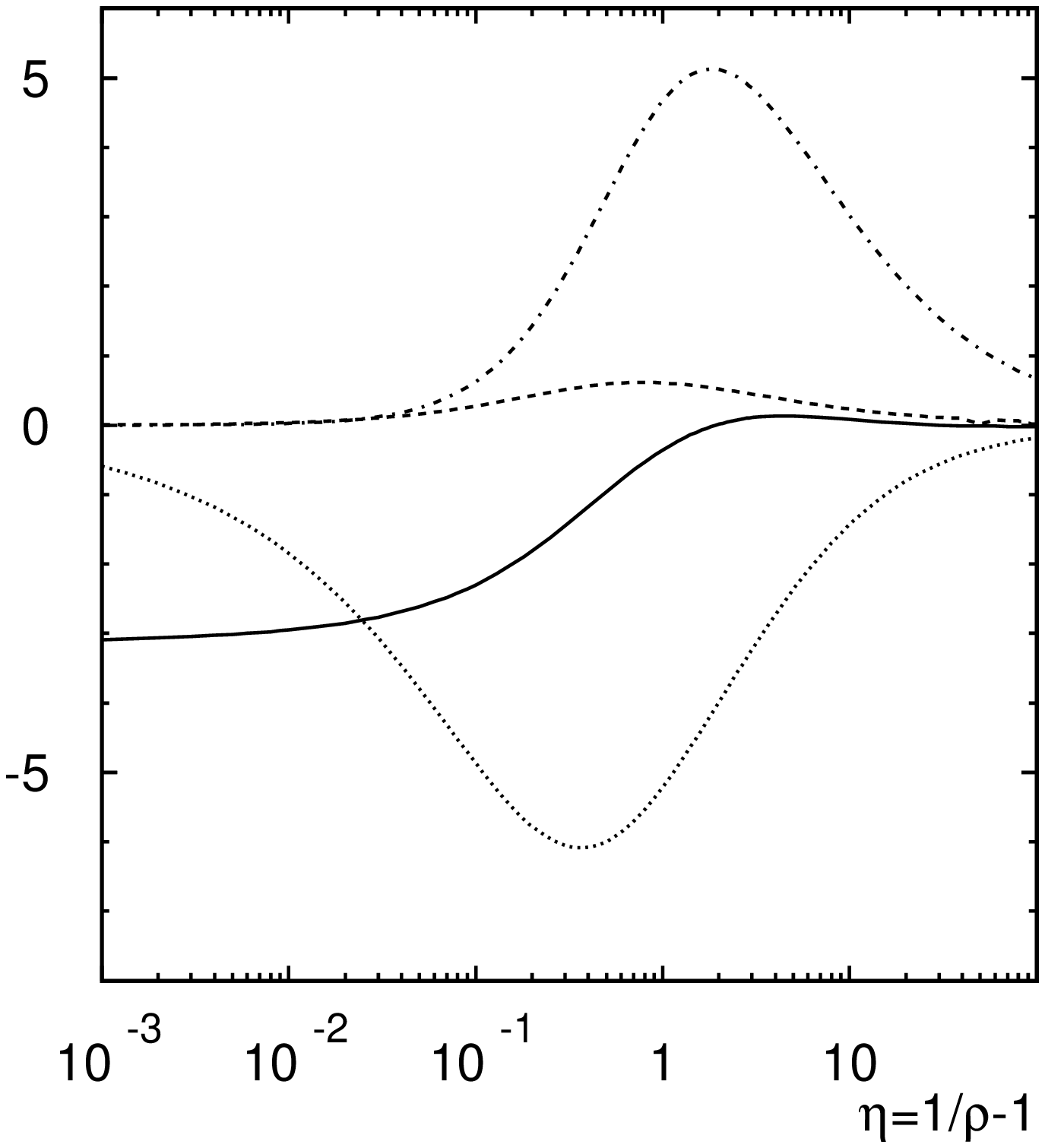,width=9cm}}
    \end{picture}
    \vskip 3.25cm
    \caption{
\it Left: Scaling functions $d^{(0)}_2(\rho,1,1)$ (dotted),
  $d^{(0)}_2(\rho,1,-1)$ (dash-dotted), $d^{(1)}_2(\rho,1,1)$ (full),
  and $d^{(1)}_2(\rho,1,-1)$ (dashed).
   Right: Scaling functions $d^{(0)}_3(\rho,1,1)$ (dotted),
  $d^{(0)}_3(\rho,1,-1)$ (dash-dotted), $d^{(1)}_3(\rho,1,1)$ (full), 
  and $d^{(1)}_3(\rho,1,-1)$ (dashed).
}
    \label{fig:o23}
  \end{center}
\end{figure}
\begin{figure}[h!]
  \unitlength1.0cm
  \begin{center}
    \begin{picture}(4.5,4.5)
      \put(-6.65,-3){\psfig{figure=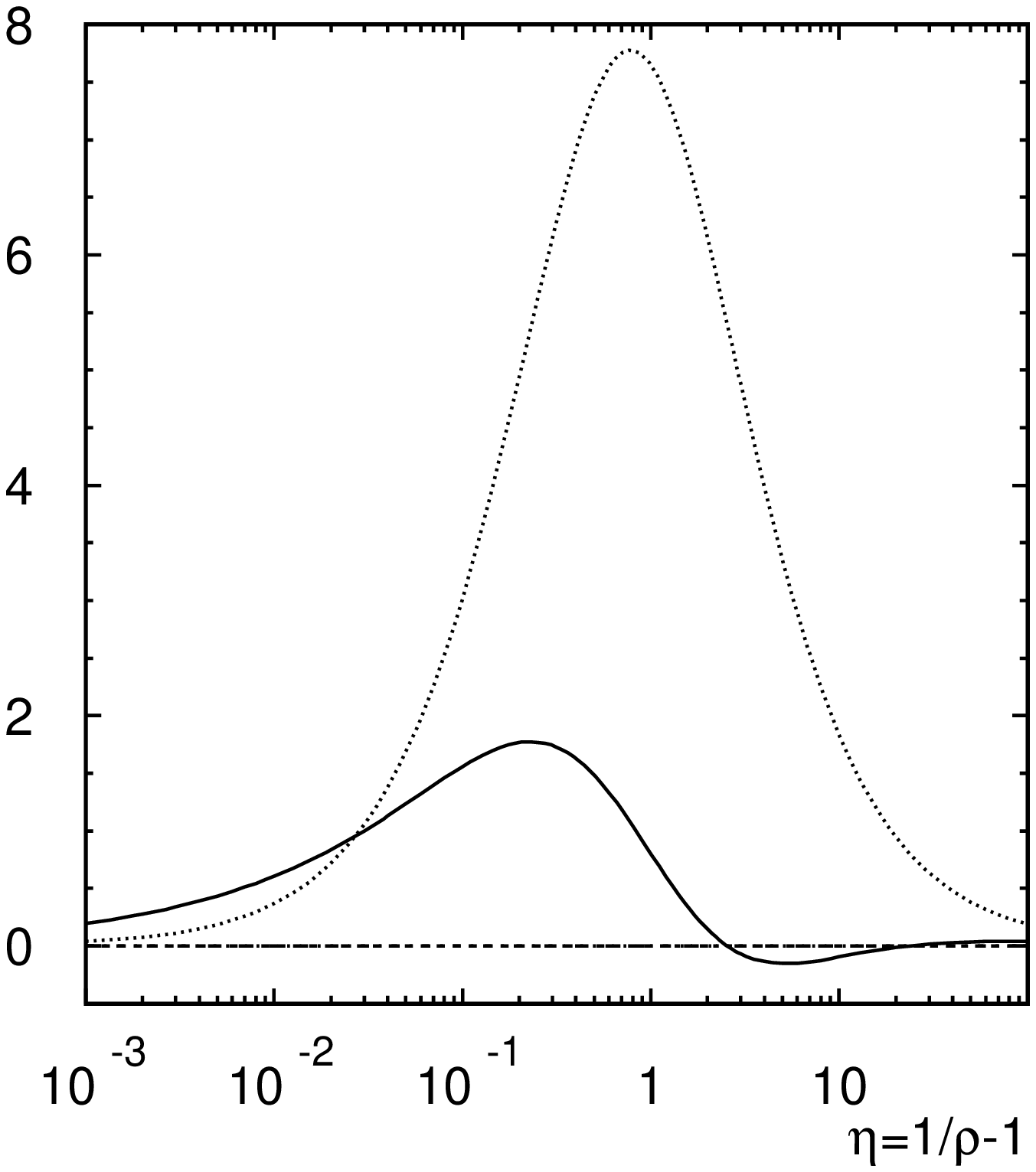,width=9cm}}
      \put(2.65,-3){\psfig{figure=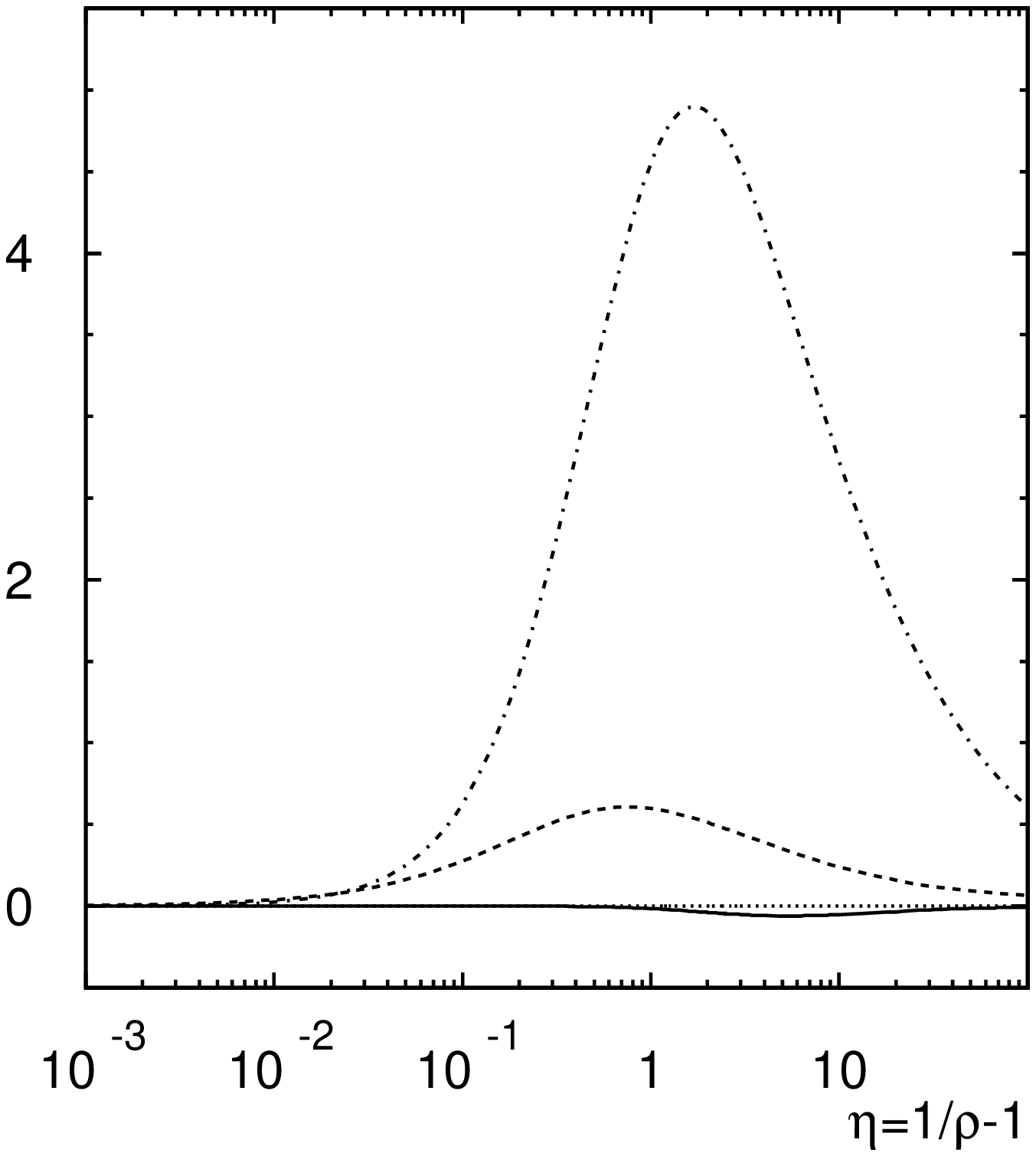,width=9cm}}
    \end{picture}
    \vskip 2.75cm
    \caption{
\it Left: Scaling functions $d^{(0)}_4(\rho,1,1)$ (dotted),
  $d^{(0)}_4(\rho,1,-1)$ (dash-dotted), $d^{(1)}_4(\rho,1,1)$ (full),
  and $d^{(1)}_4(\rho,1,-1)$ (dashed).
   Right: Scaling functions $d^{(0)}_5(\rho,1,1)$ (dotted),
  $d^{(0)}_5(\rho,1,-1)$ (dash-dotted), $d^{(1)}_5(\rho,1,1)$ (full), 
  and $d^{(1)}_5(\rho,1,-1)$ (dashed).
}
    \label{fig:o45}
  \end{center}
\end{figure}
\par
The result for unpolarized photons can be inferred from 
\begin{eqnarray}
c^{(0),(1)}(\rho,0,0)&=&{1\over 2}\left[c^{(0),(1)}(\rho,1,1)
+c^{(0),(1)}(\rho,1,-1)\right],\\
d^{(0),(1)}_{1,2,3}(\rho,0,0)&=&{1\over 2}\left[
d^{(0),(1)}_{1,2,3}(\rho,1,1)+
d^{(0),(1)}_{1,2,3}(\rho,1,-1)\right],\\
d^{(0),(1)}_{4,5}(\rho,0,0)&=&0.
\end{eqnarray}

As a check we compared our result
for the functions $c^{(0),(1)}(\rho,0,0)$ 
with the results given in Fig.~2 of Ref.~\cite{Drees:1992eh} 
and found perfect agreement. We further compared the functions  
$c^{(0),(1)}(\rho,1,\pm 1)$
to the results given in Table~1 of 
Ref.~\cite{Jikia:1996bi}. After a trivial rescaling to account for
the different conventions used in the definition of the 
scaling functions, we also found agreement.

\section{Effective cross sections and spin observables}
\label{nummerical results}
\subsection{The effective cross section for $\gamma\gamma\to t\bar{t}X$} 
The total $t\bar{t}$ cross section at a photon collider may be written 
at NLO QCD as (cf., e.g., \cite{Kuhn:1992fx})
\begin{equation}
\sigma_{t\bar{t}}=
{\alpha^2\*Q_t^4\over m^2}\int_0^{y_{\rm max}} dy_1 
\int_0^{y_{\rm max}}dy_2 f_\gamma^e(y_1,P_e,P_L) f_\gamma^e(y_2,P_e,P_L)
 \left\{
c^{(0)}+4\pi\alpha_s c^{(1)}
\right\}.
\end{equation}
The function $f_\gamma^e(y_1,P_e,P_L)$ is the normalized energy
spectrum of the photons resulting from Compton backscattering of laser light 
off the high energy electron beam. 
It is explicitly given by:
\begin{equation}
f_\gamma^e(y,P_e,P_L)= {\cal N}^{-1}\left[
{1\over 1-y}-y+(2r-1)^2-P_eP_Lxr(2r
-1)(2-y)
\right].
\label{compton}
\end{equation}
Here, $P_e(P_L)$ is the polarization of the electron (laser) beam, and
$y$ is the fraction of the electron energy in the
c.m. frame transferred to the photon. It 
takes values in the range
\begin{equation}
0 \le y \le {x\over x+1}\equiv y_{\rm max},
\end{equation}
with 
\begin{equation}
x = {4E_LE_e\over m_e^2},
\end{equation}
where $E_{L}(E_e)$ is the energy of the laser (electron) beam and $m_e$ 
is the electron mass. In order to avoid the creation of an $e^+e^-$ pair 
from the backscattered laser beam and the low energy laser beam, the maximal
value for $x$ is 
\begin{equation}
x_{\rm max} = 2(1+\sqrt{2}).
\end{equation}
For a beam energy $E_e=250$ GeV, this leads to an optimal laser energy
\begin{equation}
E_L\approx 1.26 {\rm \ eV},
\end{equation} 
which will be used in the following numerical results.
Finally,
\begin{equation}
r={y\over x(1-y)}.
\end{equation}
The normalization factor ${\cal N}$ in Eq.~(\ref{compton}) is determined by
\begin{equation}
\int_0^{y_{\rm max}}f_\gamma^e(y,P_e,P_L) dy =1.
\end{equation}
The scaling functions $c^{(0),(1)}$  have to be evaluated
at $\rho=4m^2/(y_1y_2s_{ee})$ and for polarizations
\begin{equation}
\lambda_{i}=P_\gamma(y_{i},P_{e}^{(i)},P_{L}^{(i)}),\qquad i=1,2.
\end{equation}
The function $P_\gamma(y,P_{e},P_{L})$ describes the degree of polarization
of photons scattered
with energy fraction $y$, which is given by
\begin{equation}
P_\gamma(y,P_{e},P_{L})={1\over {\cal N}\*f_\gamma^e(y,P_e,P_L)}
\left\{
xrP_e\left[1+(1-y)(2r-1)^2\right]
-(2r-1)P_L\left[
{1\over 1-y}+1-y
\right]
\right\}.
\end{equation}
Numerical results for $\sigma_{t\bar{t}}$ are given in Table~\ref{tab:sigma} for 
$\sqrt{s_{ee}}=500$ GeV and 
different polarizations of the laser and electron beam. We use the 
values $m_t=178$ GeV, $\alpha=1/128$ and $\alpha_s(\mu=m_t)=0.1$.   
\begin{table}[htbp!]
\caption{\it Results for the effective cross section at  $\sqrt{s_{ee}}=500$ GeV.}\label{tab:sigma}
\begin{center}
\renewcommand{\arraystretch}{1.3}
\begin{tabular}{|c|c|c|c|} \hline  
$(P_{e1},P_{e2};P_{L1},P_{L2})$ & $\sigma_{t\bar{t}}^{\rm LO}$ [fb] &
$\sigma_{t\bar{t}}^{\rm NLO}$ [fb] & $K=\sigma_{t\bar{t}}^{\rm NLO}/
\sigma_{t\bar{t}}^{\rm LO}$ \\ \hline
$(0,0;0,0)$          & 49.81  & 76.44  & 1.53 \\  
$(0.85,0.85;-1,-1)$  & 175.86 & 260.77 & 1.48 \\
$(0.85,0.85;+1,+1)$  & 15.96  & 26.89  & 1.68 \\
$(0.85,-0.85;-1,+1)$ & 48.99  & 71.93  & 1.47 \\ 
\hline
\end{tabular}
\end{center}
\end{table}
The QCD corrections to $\sigma_{t\bar{t}}$ are quite large. 
This is because for  $\sqrt{s_{ee}}=500$ GeV most of 
the top quark pairs are produced close to threshold where the Coulombic  
$\beta^{-1}$ singularity from soft gluons is important.
\subsection{Spin observables}
The spin observables for $\gamma\gamma\to t\bar{t}X$ discussed in section
\ref{parton level} translate into observables built from the momenta 
of the $t\bar{t}$ decay products. 
\par
The single spin asymmetries (\ref{os})
cause a nontrivial one-particle inclusive decay distribution of the form
 \begin{eqnarray}\label{distsingle}
{1\over \sigma}{d\sigma (\gamma\gamma \to a_1+X)
    \over d\cos\theta_1}=
  {1\over 2} (1 + {\rm B}_i\cos\theta_1).
\end{eqnarray}
Here, $\theta_1$ is the angle between the direction
of a top quark decay product $a_1$
measured in the top quark rest frame and one of the reference
directions $\hat{\bf a}$ defined in Eq.~(\ref{adir}).
The coefficient ${\rm B}_i$, with $i=\mbox{heli}, \mbox{beam}$ 
for  the helicity and beam bases, is determined by the top quark
spin asymmetry (\ref{os}) and by the so-called spin analysing
power of the decay product $a_1$, which will be discussed below.
\par 
The double spin asymmetries (\ref{od}) lead to a two-particle inclusive 
decay distribution of the following form:
\begin{eqnarray}
  {1\over \sigma}{d\sigma (\gamma\gamma \to a_1 a_2+X)
    \over d\cos\theta_1 d\cos\theta_2} =
  {1\over 4} (1 + {\rm B}_i\cos\theta_1+\bar{{\rm B}}_i\cos\theta_2
- {\rm C}_i \cos\theta_1 \cos\theta_2)\,\, ,
  \label{distdouble}
\end{eqnarray}
where $\theta_1$ is defined as above and $\theta_2$ is analogously the angle
between one of the top antiquark decay products and one of the 
reference directions
$\hat{\bf b}$ defined in Eq.~(\ref{abdir}). The coefficients ${\rm C}_i$
are determined by the double spin asymmetries (\ref{od}) and the spin
analysing powers of the two decay products $a_1$ and $a_2$. 
Finally, a non-zero expectation value of the observable defined in 
Eq.~(\ref{orot}) leads to a distribution of the form
\begin{equation}
{1\over \sigma}{d\sigma  (\gamma\gamma \to a_1 a_2+X)\over d\cos\varphi}=
{1\over 2} (1 - {\rm D} \cos\varphi),
\label{distrot}
\end{equation}
where $\varphi$ is the angle between the direction of flight of the top
decay product $a_1$ and the antitop decay product $a_2$ defined in the $t$ 
and $\bar{t}$ rest frames, respectively. We recall that these rest frames
have to be obtained by a rotation-free boost from the $t\bar{t}$-ZMF.
\par
The spin analysing power of the $t$ and $\bar{t}$ decay products is 
encoded in  
the one-particle inclusive angular distributions
$d\Gamma/d\cos\theta$ for the decays
\begin{eqnarray} 
  t(s_t) \to a_1(q_1) + X_1 \, ,\nonumber \\
  {\bar t}(s_{\bar t}) \to a_2(q_2) + X_2 \, .
\end{eqnarray}
Here  $q_1$ and $q_2$
are the momenta  of $a_1$ and $a_2$, respectively, 
defined in the rest frame
of the (anti)top quark. and $\theta$
is the angle between the polarization vector of the (anti)top quark
and the direction of flight of $a_1(a_2)$. 
For a fully polarized ensemble of top
quarks (antiquarks) these distributions are of the form 
\begin{equation}
 {d\Gamma^{(1,2)}\over d\cos\theta}
={\Gamma^{(1,2)}\over 2}(1\pm\kappa^{(1,2)} \cos\theta) \, , 
\label{andist}
\end{equation}
where $\Gamma^{(1,2)}$ is the partial width of the
respective decay channel. The quantity 
$\kappa^{(1,2)}$ is the (anti)top-spin 
analysing power of $a_{1,2}$. For the case of the
standard $(V-A)$ charged current interactions these distributions
were computed to order $\alpha_s$ for the semileptonic and
non-leptonic channels in Refs.~\cite{Czarnecki:1990pe}
and \cite{Brandenburg:2002xr}, respectively. 

As we work to lowest order in the electroweak couplings,
${\Gamma^{(2)}} = \Gamma^{(1)}$ and 
${\kappa}_2 = {\kappa}_1$ to all orders in $\alpha_s$, if the channel
$a_2 + X_2$ is the
charge-conjugate of $a_1 + X_1$. 
\par
For  semileptonic top
decays $t \to b \ell^+{\nu}_{\ell} (g)$, 
the charged lepton is the most efficient analyser of the spin
of the top quark.
In the case of  non-leptonic  decays $t \to b q {\bar q}'(g)$ 
a good top spin analyser that can be identified easily is the 
least-energetic light quark jet.
\par
In Ref.~\cite{Brandenburg:2002xr}  
the coefficients $\kappa^{(f)}$ were given to  NLO 
accuracy for different choices of the spin analyser. 
To compute the coefficients ${\rm B}_i,\ {\rm C}_i$ and ${\rm D}$
we  need  the partial widths
\begin{equation}
\Gamma^{({\rm sl,h})} = a^{({\rm sl,h})}_0 + 4\*\pi\*\alpha_s \*a^{({\rm sl,h})}_1,
\label{expga}
\end{equation}
where the indices ${\rm sl}$ and ${\rm h}$ stand for semileptonic and hadronic
decay modes. Further, we need
the dimensionful coefficients
\begin{equation}
 \Gamma^{({\rm sl,h})} \kappa^{(\ell,j)} =
b^{(\ell,j)}_0 +4\*\pi\*\alpha_s\* b^{(\ell,j)}_1 \, ,
\label{expka}
\end{equation}
where $\ell$ ($j$) refers to using the charged lepton (least-energetic light 
quark jet) as spin analyser. 
For the determination of these coefficients we use the Fermi constant 
$G_F=1.16639\times 10^{-5}\mbox{ GeV}^{-2}$,
$m$ = 178 GeV, $m_W$ = 80.42 GeV,  $\Gamma_W$ = 2.12 GeV,
$m_b$ = 4.75 GeV, and all other quark and lepton masses are put to zero.
(We do not use the narrow width
approximation for the intermediate  $W$ boson.)
We obtain, 
putting the CKM matrix elements $|V_{tb}|=|V_{qq'}|=1$: 
\begin{eqnarray}
a_0^{\rm h}  &=& 0.52221 {\rm \ GeV}, \nn \\ 
a_0^{\rm sl} &=&  {a_0^{\rm h} \over N},\nn \\
a_1^{\rm h}  &=& -0.01968(15)  {\rm \ GeV}, \nn \\
a_1^{\rm sl} &=& -0.01097(5)   {\rm \ GeV}.
\end{eqnarray}
For the  relevant coefficients $b_{0,1}$
we obtain:
\begin{eqnarray}
b_0^{\ell}&=& a_0^{\rm sl}, \nn \\
b_0^{j}&=&  0.26950 {\rm \ GeV},\nn \\
b_1^{\ell}&=& -0.01118(8) {\rm \ GeV}, \nn \\
b_1^{j}&=&  -0.02375(26) {\rm \ GeV}.
\end{eqnarray}
The Durham algorithm was used as jet clustering scheme 
to obtain the four parton contribution to $b_1^{j}$.
\par
Within the leading pole approximation for the intermediate 
top quarks and antiquarks, the coefficients 
of the single and double differential distributions
(\ref{distsingle})--(\ref{distrot}) are obtained in terms
of the following quantities:
\begin{eqnarray}
\sigma_s&=& {\alpha^2\*Q_t^4\over m^2}{1\over \Gamma_t}\int_0^{y_{\rm max}} dy_1 
\int_0^{y_{\rm max}}dy_2 f_\gamma^e(y_1,P_e,P_L) f_\gamma^e(y_2,P_e,P_L)
\nonumber \\ &\times& \left\{
c^{(0)}a_0^{(1)}+4\pi\alpha_s\left[
c^{(1)}a_0^{(1)}+c^{(0)}a_1^{(1)}
\right]
\right\}, \\
\sigma_d&=& {\alpha^2\*Q_t^4\over m^2}
{1\over \Gamma_t^2}\int_0^{y_{\rm max}} dy_1 
\int_0^{y_{\rm max}}dy_2 f_\gamma^e(y_1,P_e,P_L) f_\gamma^e(y_2,P_e,P_L)
\nonumber \\ &\times& \left\{
c^{(0)}a_0^{(1)}a_0^{(2)}+4\pi\alpha_s\left[
c^{(1)}a_0^{(1)}a_0^{(2)}+c^{(0)}a_1^{(1)}a_0^{(2)}
+c^{(0)}a_0^{(1)}a_1^{(2)}
\right]
\right\}, \\
N_{r}^s&=& {\alpha^2\*Q_t^4\over m^2}
{1\over \Gamma_t}\int_0^{y_{\rm max}} dy_1 
\int_0^{y_{\rm max}}dy_2 f_\gamma^e(y_1,P_e,P_L) f_\gamma^e(y_2,P_e,P_L)
\nonumber \\ &\times&\left\{
d^{(0)}_rb_0^{(1)}+4\pi\alpha_s\left[
d^{(1)}_rb_0^{(1)}+d^{(0)}_rb_1^{(1)}
\right]
\right\}, \\
N_{r}^d&=& {\alpha^2\*Q_t^4\over m^2}
{1\over \Gamma_t^2}\int_0^{y_{\rm max}} dy_1 
\int_0^{y_{\rm max}}dy_2 f_\gamma^e(y_1,P_e,P_L) f_\gamma^e(y_2,P_e,P_L)
\nonumber \\ &\times&\left\{
d^{(0)}_rb_0^{(1)}b_0^{(2)}+4\pi\alpha_s\left[
d^{(1)}_rb_0^{(1)}b_0^{(2)}+d^{(0)}_rb_1^{(1)}b_0^{(2)}
+d^{(0)}_rb_0^{(1)}b_1^{(2)}
\right]
\right\}.
\end{eqnarray}
We then get to NLO in $\alpha_s$:
\begin{equation}
{\rm D}={N_1^d\over \sigma_d},\qquad {\rm C}_{\rm heli}={N_2^d\over \sigma_d},
\qquad {\rm C}_{\rm beam}={N_3^d\over \sigma_d},\qquad 
{\rm B}_{\rm heli}={N_4^s\over \sigma_s},
\qquad {\rm B}_{\rm beam}={N_5^s\over \sigma_s}.
\end{equation}
The LO and NLO results for these quantities are shown in Table~\ref{tab:dcb}
for favorable and realistic 
choices of electron and laser polarizations, using the same
parameters as in Table~\ref{tab:sigma}.
In most cases the QCD corrections are of the order of a few percent
and thus much smaller than the corrections to the total $t\bar{t}$ cross
section. This was to be expected, since the bulk of the corrections
is due to soft gluons which do not affect the $t\bar{t}$ spin state. 
The biggest correction ($\sim 11 \%$)  
occurs for the  coefficient ${\rm B}_{\rm beam}$ 
if the least energetic light jet
is used as spin analyser. 
\par
Photon polarization is an important asset: It is necessary to 
obtain polarized top quarks and thus non-zero coefficients  
${\rm B}_{\rm heli }$ and ${\rm B}_{\rm beam}$. Further, the choice
$(P_{e1},P_{e2};P_{L1},P_{L2})=(0.85,0.85;-1,-1)$, which increases the 
total yield of $t\bar{t}$ pairs by more than a factor of 3 (see Table~1), 
in addition leads to larger $t\bar{t}$ spin correlations. 
In particular, in the 
helicity basis the correlation is then almost 100\% in the dilepton channel.
\par
\begin{table}[htbp!]
\caption{\it Results for double and single spin asymmetries
 at $\sqrt{s_{ee}}=500$ GeV.}\label{tab:dcb}
\begin{center}
\renewcommand{\arraystretch}{1.3}
\begin{tabular}{|c|c|cc|cc|cc|} \hline  
  &  &  \multicolumn{2}{|c|}{dilepton}  
& \multicolumn{2}{|c|}{lepton-jet}  &  \multicolumn{2}{|c|}{jet-jet}  
\\ 
 & $(P_{e1},P_{e2};P_{L1},P_{L2})$ & LO & NLO & LO & NLO & LO & NLO \\ \hline
${\rm D}$  & $(0,0;0,0)$ & $-0.670$ & $-0.686$ &$-0.346$ &$-0.338$ &$-0.178$ & $-0.167$ \\ 
 & $(0.85,0.85;-1,-1)$ & $-0.806$ & $-0.801$ & $-0.416$ &$-0.394$ &$-0.215$ & $-0.194$\\
\hline
 ${\rm C}_{\rm heli}$ & $(0,0;0,0)$ & $\phantom{-}0.811$ &$\phantom{-}0.826$ 
&$\phantom{-}0.418$ &$\phantom{-}0.408$ &$\phantom{-}0.216$ & $\phantom{-}0.201$ \\ 
 & $(0.85,0.85;-1,-1)$ & $\phantom{-}0.985$ & $\phantom{-}0.981$& 
$\phantom{-}0.508$ &$\phantom{-}0.483$ &$\phantom{-}0.262$ &$\phantom{-}0.238$ \\
\hline
 ${\rm C}_{\rm beam}$ & $(0,0;0,0)$ & $-0.580$ &$-0.606$ &$-0.299$ & $-0.299$&$-0.154$
 & $-0.148$ \\ 
 & $(0.85,0.85;-1,-1)$ & $-0.808$ &$-0.804$ &$-0.417$ & $-0.396$& $-0.215$ & $-0.195$\\
\hline \hline
  &  &  \multicolumn{2}{|c|}{lepton+X}  
& \multicolumn{2}{|c|}{jet+X}  &   \multicolumn{2}{|c}{}
\\ 
 & $(P_{e1},P_{e2};P_{L1},P_{L2})$ & LO & NLO & LO & NLO &
\multicolumn{2}{|c}{}  \\ \cline{1-6}
${\rm B}_{\rm heli}$ 
 & $(0.85,0.85;-1,-1)$ &$\phantom{-}0.658$ &$\phantom{-}0.655$ &
$\phantom{-}0.340$ &$\phantom{-}0.323$ & \multicolumn{2}{|c}{}
\\  \cline{1-6}
${\rm B}_{\rm beam}$ 
 & $(0.85,-0.85;-1,1)$ & $-0.684$  &$-0.637$ &$-0.353$ &$-0.314$ &
 \\  \cline{1-6}
\end{tabular}
\end{center}
\end{table}
So-called non-factorizable corrections do neither contribute at NLO QCD
to $\sigma_{t\bar{t}}$ nor to the angular correlations considered
above. A proof of this statement is given in \cite{Bernreuther:2004jv}.


\section{Conclusions}
We have computed a variety of  spin observables for the process 
$\gamma\gamma\rightarrow t\bar{t} X$ up to order 
$\alpha^2 \alpha_s$. Together with the differential rates of polarized
top and antitop quark decays at order $\alpha_s$, 
we have obtained the NLO QCD contributions to 
the fully differential cross section with intermediate top quark pair 
production at a photon collider.

We have applied the above results to 
$t\bar{t}$ production and decay
at a future linear collider operating at $\sqrt{s}=500$ GeV. 
We have shown that for an appropriate choice of 
the polarizations of the laser and electron beam,
the cross section and the double/single spin asymmetries
can be quite large. 
While the QCD corrections to 
the cross section can be very large,  
most of the double/single spin asymmetries are affected
at the level of only a few percent.
The observables considered here will provide useful tools 
to analyse in detail the
top quark pair production and decay dynamics. In particular,
their precise measurement will test whether the top quark
truly behaves as a quasi-free fermion as predicted in the Standard Model.

\section*{Acknowledgments}
We wish to thank W.~Bernreuther and P.~M.~Zerwas for discussions. 
A.B. was supported by a Heisenberg grant of the Deutsche
Forschungsgemeinschaft. Z.G.~Si wishes 
to thank DFG and MoE of China for financial support of his
short visit in Germany, and also thanks DESY Theory Group for its hospitality.

\end{document}